\documentclass[%
reprint,
amsmath,
amssymb,
prx,
aps,
nobalancelastpage,
superscriptaddress, 
]{revtex4-2}

\usepackage{graphicx,xcolor,amsfonts, microtype, braket}
\usepackage[colorlinks,allcolors=blue]{hyperref}
\usepackage{orcidlink}
\usepackage{mathtools}
\usepackage[capitalize]{cleveref}
\let\mysection\section 
\let\mysubsection\subsection 

\newcommand{\eps}{\varepsilon}
\newcommand{\dagg}{^\dagger}
\newcommand{\pdagg}{^{\vphantom{\dagger}}}
\renewcommand{\H}{\mathcal H}
\renewcommand{\vec}[1]{\mathbf{#1}}
\newcommand{\mat}[1]{\begin{pmatrix} #1 \end{pmatrix}}
\DeclareMathOperator{\diag}{diag}
\DeclareMathOperator{\tr}{Tr}

\newcommand{\K}{\mathcal K}

\begin{document}
\title{Gaussian tomography for cold-atom simulators}

\author{Matthew Kiser\orcidlink{0000-0002-9357-7583}}
\thanks{Part of this work was done while at Volkswagen AG}
\email{matthew.kiser@tum.de}
\affiliation{TUM School of Natural Sciences, Technical University of Munich, Boltzmannstr. 10, 85748 Garching, Germany}
\affiliation{IQM Quantum Computers, Georg-Brauchle-Ring 23-25, 80992, Munich, Germany}

\author{Max McGinley\orcidlink{0000-0003-3122-2207}}
\affiliation{T.C.M. Group, Cavendish Laboratory, JJ Thomson Avenue, Cambridge CB3 0HE, United Kingdom}

\author{Daniel Malz\orcidlink{0000-0002-8832-0927}}
\affiliation{Department of Mathematical Sciences, University of Copenhagen, 2100 Copenhagen, Denmark}

\date{\today}

\begin{abstract}
	
	A limitation of analog quantum simulators based on cold atoms in optical lattices is that readout is typically limited to observables diagonal in the charge basis, i.e., densities and density correlation functions.
	To overcome this limitation, we propose experiment-friendly schemes to measure charge-off-diagonal correlations (such as currents).
	Our protocols use non-interacting dynamics for random times followed by standard quantum gas microscope measurements to effectively measure in random bases.
	The main requirement of our scheme is the ability to turn off interactions, which can be done in many atomic species using Feshbach resonances. Importantly, our scheme requires no local control and otherwise also exhibits modest requirements in terms of total evolution time and number of repetitions. 
	We numerically demonstrate efficient estimation of bilinear correlation functions, requiring less than $4000$ samples to measure local currents to 5\% error (system-size independent) and $\sim 10^4$ samples to simultaneously measure all non-local correlations in 70-site systems. Due to its simplicity, our protocol is implementable in existing platforms and thus paves the way to precision measurements beyond particle number measurements.
\end{abstract}

\maketitle
\mysection{Introduction}Analog quantum simulators are table-top experiments designed to study complex many-body physics and quantum dynamics~\cite{feynman1982simulating,lloyd1996universal,cirac2012goals, georgescu2014quantum, altman2021quantum, trivedi2024quantum}, such as those described by the Hubbard model \cite{bloch2008manybody, esslinger2010fermihubbard,tarruell2018quantum,arovas2022hubbard,xu2025neutral}. A critical component of these experiments is the post-evolution measurement step to determine physical quantities of interests such as currents, correlation functions, or the system's energy.

Optical lattices with cold atoms are a leading platform for analog simulation, capable of simulating systems of hundreds of strongly interacting particles that mimic the behavior of strongly correlated electrons~\cite{gross2017quantum, browaeys2020manybody,altman2021quantum, daley2022practical}.
A key breakthrough was the quantum gas microscope, which allows single site-resolved measurement of each particle at the end of the experiment~\cite{Bakr2009,sherson2010singleatomresolved,endres2011observataion,haller2015singleatom,kuhr2016quantum,gross2021quantum, su2025fast}.
The main limitation of the quantum gas microscope is that it only allows measurements in the particle number (or computational) basis, which excludes important observables such as currents.

There is a significant effort to overcome these limitations.
Experimentally, nearest-neighbor current measurements have recently been achieved by engineering a superlattice~\cite{impertro2024local}, which constitutes a step towards realizing local control.
While promising, extending this approach to longer-range correlations appears challenging.  More recently, several proposals to access more general observables have been put forward that use the natural evolution of the quantum simulator to effectively change the measurement basis~\cite{Ohliger2013,Elben2018,vermersch2018unitary,vermersch2019probing,brydges2019probing,gluza2020quantum, joshi2020quantum,elben2020manybody,gluza2021recovering,VanKirk2022,tran2023measuring, mcginley2023shadow, hu2023classical,notarnicola2023randomized,naldesi2023fermionic,denzler2024learning, liu2024predicting, Mark2025}.
Conceptually, such approaches bear resemblance to time-of-flight measurements, where the momentum distribution of a tightly confined gas is mapped through free evolution onto a measurable spatial distribution~\cite{Anderson1995,Davis1995}.

Inspired by randomized and shadow tomography schemes designed for digital quantum computers \cite{aaronson2018shadow,huang2020predicting,zhao2021fermionic,elben2022randomized, wan2023matchgate,hu2025demonstration, heyraud2025unified}, these proposals employ time evolution under an interacting~\cite{Ohliger2013,Elben2018,vermersch2018unitary,joshi2020quantum,elben2020manybody,VanKirk2022,tran2023measuring, mcginley2023shadow,hu2023classical} or non-interacting~\cite{gluza2020quantum,gluza2021recovering,naldesi2023fermionic, denzler2024learning,liu2024predicting,Mark2025} Hamiltonian for random times or in the presence of ancillas, before measuring. When the ensemble of times is chosen correctly, the net effect is a measurement in a random basis, which allows one to reconstruct many observables by classically post-processing the measured data. However, schemes using interacting dynamics require either local control or the ability to simulate long-time evolution during the classical post-processing, which we assume here to be prohibitive. Approaches that use non-interacting dynamics to randomize measurement bases are more scalable, because such dynamics are easy to simulate classically~\cite{gluza2020quantum,gluza2021recovering,naldesi2023fermionic,denzler2024learning,Mark2025}.
Closest to the present work is Ref.~\cite{gluza2021recovering}, especially in motivation. There it was shown that it is possible to fit a covariance matrix to measurement data obtained from non-interacting quenches, but the algorithm is different, and it is not clear how it scales, or whether it admits local recovery. 
Also relatedly, Ref.~\cite{denzler2024learning} has proved rigorously that even the time evolution produced by an ensemble of random translation-invariant hopping Hamiltonians is sufficient to recover all two-point and four-point correlation functions (albeit in times that scale as a quartic of system size), which is prohibitive for current experiments, which routinely exceed $100$ sites.

Here, we propose and analyze schemes that achieve simple and efficient estimation of correlation functions and thus overcome the aforementioned challenges.
Specifically, we consider lattice-based cold-atom quantum simulators with three main ingredients: (i) the ability to turn off interactions, leading to non-interacting dynamics, (ii) the ability to measure the density distribution of the atoms, and (iii) an additional laser used to impart a quasiperiodic potential that breaks all lattice symmetries (see \cref{fig:overview_protocols}). With these ingredients, our scheme allows one to simultaneously measure all local correlation functions to high accuracy, using time evolution for at most a few hopping times and a constant number of repetitions (sample complexity) of the order of less than $10^4$. Measuring arbitrary non-local correlation functions is substantially harder, but we present a protocol in which a 1d chain of length $N$ expands into a 2d $N\times N$ lattice (reminiscent of time-of-flight measurements), in which hopping times of order $N$ are sufficient to measure all correlation functions.

\begin{table}[t]
    \begin{ruledtabular}
        \begin{tabular}{ll}
            \textbf{Local scheme} & \textbf{Global scheme}\\
            \hline
            {\includegraphics[width=0.18\textwidth]{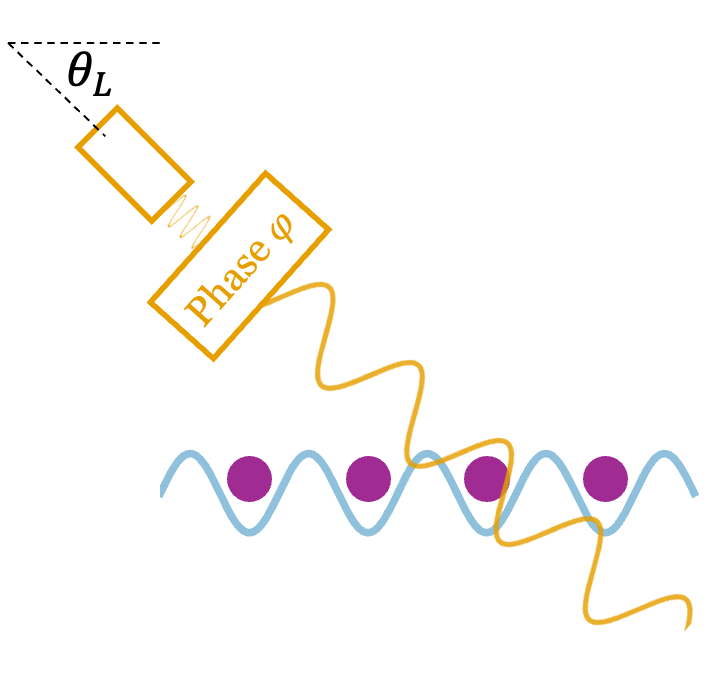}} &
            {\includegraphics[width=0.18\textwidth]{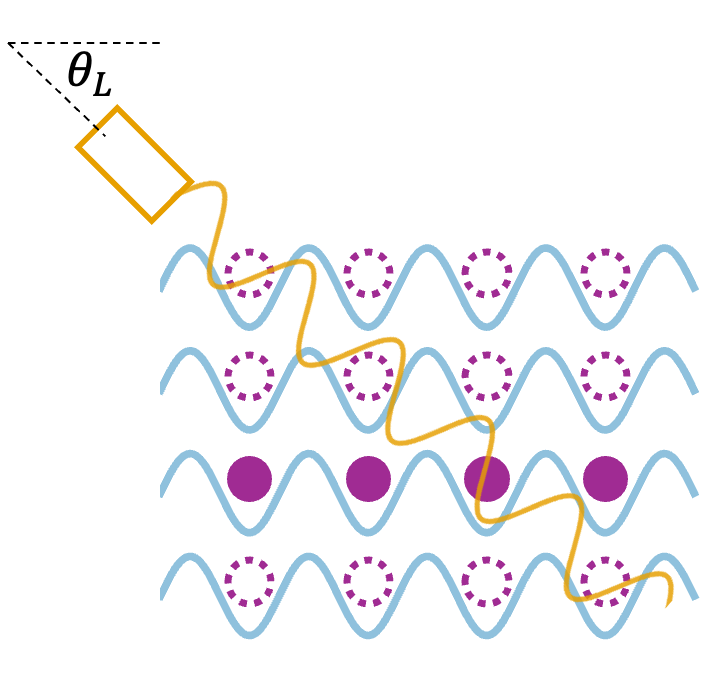}}\\
            Atoms hopping in 1d or 2d & Expansion from 1d$\to$2d \\
            $\to$ No ancillas & $\to$ $\Omega(N^2)$ ancillas \\
            Random times $t_i\leq $ const & Fixed $t\propto N$ \\
            Applied quasiperiod pot.\ (orange) & Same \\
            Random pot.\ strength $h_i\leq h_\mathrm{max}$ & Same \\
            Random pot.\ phase $\varphi_i$ & No phase \\
        \end{tabular}
    \end{ruledtabular}
    \caption{Illustration and main characteristics of the two schemes that we simulate numerically.
    In the local scheme, we only aim to estimate local bilinear observables. In the global scheme we estimate the whole correlation matrix.}
\label{fig:overview_protocols}
\end{table}

\mysection{Setting}\label{sec:setting}We consider experiments based on spinless fermionic atoms hopping in an optical lattice, but various generalizations, e.g., to non-interacting spinful fermions or to bosons, are straightforward.
We assume that we can repeatedly prepare the same state $\rho_0$ (the outcome of the experiment) and would like to measure its correlations, which are captured by the correlation matrix
\begin{equation}
	C_{0,\vec i\vec j} = \text{Tr}[\rho_0 c_{\vec i}^\dagger c_{\vec j}]
\end{equation}
where $ c_{\vec i}$ is the annihilation operator for the fermion on site $\vec i$, with $\{c_{\vec i}^{ },c_{\vec j}\dagg\}=\delta_{{\vec i},{\vec j}}$. For simplicity and efficiency, we focus here on bilinear observables (i.e., the correlation matrix), but our method readily generalizes to higher-order correlations at the expense of an increased sample complexity, as shown in \cref{app:higher_k_values}. We consider both 1d and 2d arrays, and in the latter case $\vec i=(i_x,i_y)$.

The key obstacle to measuring the correlation matrix is that read-out in typical cold--atom experiments is based on quantum gas microscopes, which are based on detecting the atoms and thus allow measurements only in the number basis.
To cirumvent this limitation, we propose evolving $\rho_0$ under the non-interacting Hamiltonian of the simulator
\begin{equation}\label{eq:hop_ham_ferm_used}
	\begin{aligned}
		H(h,\phi) &= \vec c\dagg\cdot\H(h,\phi)\cdot\vec c
		\\&=-\sum_{\langle\vec i\vec j\rangle} c_{\vec i}\dagg c_{\vec j}^{ } + h\sum_{\vec i}\cos({\vec k}\cdot{\vec i}+\phi)c_{\vec i}^\dagger c_{\vec i},
	\end{aligned}
\end{equation}
where $\vec c$ is a column vector containing the system annihilation operators $c_\vec{i}$.
The hopping part of this Hamiltonian describes typical cold atom experiments based on optical lattices. The second term is a quasiperiodic potential, which could be realised by either a second laser at a different wavelength, or by re-using the lattice laser, but shining it at at an angle $\theta_{\rm L}$, giving $\vec k=2\pi(\cos\theta_{\rm L},\sin\theta_{\rm L})$.
The additional potential is crucial to break the symmetries of the lattice (\cref{sec:Minimum}).

In the following, we present a general protocol, but will select two concrete variants for numerical simulation, illustrated in \cref{fig:overview_protocols}, which are designed for complementary purposes:
\begin{enumerate}
	\itemsep0em
	\item[(i)] A ``local scheme'', in which we want to measure only local correlations. This requires evolution times of only a few hopping times ($<2$, and independent of system size), and we assume that the phase $\phi$ and strength $h$ of the quasiperiodic potential can be changed for each repetition of the experiment. 
	\item[(ii)] A ``global scheme'' to estimate arbitrary correlation functions. Here we consider a 1d system of length $N$ expanding into two dimensions, which requires evolution times of order $N$, with a fixed laser phase and strength.
\end{enumerate}
More variations are conceivable, and can be adapted to experimental suitability and needs.

\mysection{Methods}\label{sec:methods}We now give a precise mathematical description of our protocol and show how to compute an estimator for the correlation matrix from the measurement data.

\emph{Experimental steps.---}The first (optional) step in the protocol is to attach an $N_{\rm anc}$-site ancilla system. In practice, if for example the atoms were initially confined to move in a 1d lattice, one could lower the potential barrier and allow them to explore a 2d lattice. 
The correlation matrix of the system plus ancillas is then
\begin{equation}
	C_\mathrm{tot}=C_0\oplus C_\mathrm{anc}.
\end{equation}
Here, we consider an empty ancilla system, where the correlation matrix is the zero matrix, $C_\mathrm{anc}=0$, but the protocol can easily be generalized to ancillas in some other state with known correlation matrix.

In the second step of the protocol, we evolve many copies of the system by randomly chosen non-interacting Hamiltonians of the form in \cref{eq:hop_ham_ferm_used}.
Due to the non-interacting nature of the Hamiltonian, the final state in the simulator can be described efficiently by a unitary transformation of the initial covariance matrix
\begin{equation}
	C_{s}=U_{s}^*(C_0\oplus C_\mathrm{anc})U_{s}^T,
	\label{eq:CorrEvol}
\end{equation}
where $U_s$ are drawn according to probabilities $p_s$ from an ensemble $\mathcal U=\{(p_s, U_s=e^{-it_s\mathcal{H}(h_s, \phi_s)} )\}_{s=1}^S$ where time $t_s\in(0,t_\mathrm{max}]$, potential strength $h_s\in(0,h_\mathrm{max}]$ and phase $\phi_s\in[0,2\pi]$.

In the last experimental step, we measure the particle occupation on each site of the entire system, which yields the random variables $\hat{n}_{\vec j}$ (here and in the following, we use hats to denote random variables). These random variables can be arranged into a column vector $\hat{\vec n}$ of dimension $N_{\rm tot} = N+N_{\rm anc}$. Their expectation values (conditioned on the choice $s$) are
\begin{equation}
	\vec d_s \coloneqq \mathbb{E}[\hat{\vec n}|s]  = \diag(C_s) = F_s|C_0) + F_s^{\rm anc}|C_{\rm anc}),
	\label{eq:OccExpectation}
\end{equation}
where $|C_0)$ is a $N^2$-dimensional vector containing each matrix element of $C_0$, and similarly for $|C_{\rm anc})$. $F_s$ and $F^{\rm anc}_s$ are linear maps that are implicitly defined by restricting the matrix equation \cref{eq:CorrEvol} to diagonal elements and separating the contributions from $C_0$ and $C_\mathrm{anc}$.

A single run of the experiment thus generates a random choice of unitary labelled by $\hat{s}$ and a vector of particle numbers $\hat{\vec n} \in \{0,1\}^{N_{\rm tot}}$. We collate these data into a sub-block of a larger vector $\hat{\vec{z}} = \hat{\vec{n}} \otimes \vec e_{\hat s}$, where $\{\vec e_s\}_{s=1}^{S}$ are the standard basis vectors of dimension $S$ and $\otimes$ is the Kronecker product. Over both the random choice $\hat{s}$ and the measurement outcomes $\hat{\vec n}$, the expectation value is
\begin{align}
	\mathbb{E}[\hat{\vec z}] = (p_1\vec d_1 \; \cdots \; p_S \vec{d}_S)^T = F|C_0) + F^{\rm anc}|C_{\rm anc})
    \label{eq:z-expect}
\end{align}
where $F = (p_1F_1 \quad p_2F_2 \quad \cdots \quad p_SF_S)^T$, and similar for $F^{\rm anc}$. If $F$ has full row rank, we can find a left inverse $G$ such that $GF = I$. Thanks to linearity, the random variable $|\hat{Y}) \coloneqq G(\vec z - \vec d_{\rm anc})$, where $\vec d_{\rm anc} = F^{\rm anc}|C_{\rm anc})$, will constitute an unbiased estimator of $|C_0)$, i.e.~its expectation value is the desired correlation matrix $\mathbb{E}[|\hat Y)] = |C_0)$. 

\emph{Postprocessing.---}Running the experiment $R$ times, with a different randomly chosen $s$ in each repetition, we generate multiple pairs of data $\{(\hat{s}_r, \hat{\vec n}_r)\}_{r=1}^R \equiv \{\hat{\vec z}_r\}_{r=1}^R$, each of which can be used to construct an independent estimator of the full correlation matrix $|\hat{Y}_r)$ as described above. The sample mean $|\hat{Y}_{\rm mean}) = \frac{1}{R}\sum_{r=1}^R |\hat{Y}_r)$ will then tend towards $|C_0)$ for sufficiently large $R$. To obtain an estimate $\hat{\theta}_O$ for the expectation value $\text{Tr}[\rho_0 O]$ of a particular quadratic observable $O = \sum_{\vec i \vec j}o_{\vec i \vec j} c^\dagger_{\vec i} c_{\vec j}$, we can compute
\begin{align}
	\hat{\theta}_O \coloneqq (o|\hat{Y}_{\rm mean}) = \frac{1}{R}\sum_{r = 1}^R (o| \cdot G \cdot \hat{\vec z}_r - \mu_O
	\label{eq:EstimatorA}
\end{align}
where $\mu_O = (o| \cdot G \cdot \vec d_{\rm anc}$ is a non-random variable. By construction, the above estimator satisfies $\mathbb{E}[\hat{\theta}_O] = \text{Tr}[\rho_0 O]$. As in shadow tomography \cite{elben2022randomized,aaronson2018shadow,huang2020predicting}, the choice of observable only appears in the post-processing; thus, we can simultaneously estimate many correlations from the same set of experimental data.

Since $\hat{\theta}_O$ is a random variable, we expect some statistical error $\Delta_O = |\hat{\theta}_O - \text{Tr}[\rho_0 O]|$. The required number of repetitions (the `sample complexity') to achieve a desired level of statistical uncertainty can be determined by calculating the variance of the estimator \cref{eq:EstimatorA}. We find an approximate expression
\begin{align}
	\text{Var}[\hat{\theta}_O] \approx \frac{1}{R} (o|G W G^\dagger |o)\,,
	\label{eq:VarianceGeneral}
\end{align}
where $W = \bigoplus_{s=1}^S p_s W_s$ is a block-diagonal matrix that depends on the ensemble and the ancilla state, which are both known beforehand. See \cref{app:variance} for an explicit expression for $W$, and a proof of the above.

Since the left inverse $G$ may not be unique when $SN_{\rm tot} > N^2$, we wish to choose the estimator that minimizes the variance, and hence the statistical uncertainty in our estimation scheme. In \cref{app:variance} we show this to be
\begin{align}
	G &= L^{-1} F^\dagger W^{-1}, & L = F^\dagger W^{-1} F\,.
	\label{eq:OptimalInverse}
\end{align}
We observe numerically that this choice can dramatically outperform e.g.~the Moore-Penrose pseudoinverse $G = (F^\dagger F)^{-1}F^\dagger$. This approach reduces to the Moore-Penrose pseudoinverse in the regime where no ancilla sites are present and there are many unitaries in the ensemble.

\mysection{Quantifying Sample Complexity}\label{sec:sample_complexity}The practical utility of our framework depends on the sample complexity---the number of experimental samples needed to achieve a particular accuracy $\Delta_O \leq \varepsilon$---which itself is a function of both the ensemble of unitaries $\mathcal{U}$ and the observable $O$ being estimated. Using our optimal inverse \cref{eq:OptimalInverse}, the variance \cref{eq:VarianceGeneral} of the estimator is
\begin{align}
	\text{Var}[\hat{\theta}_O]_{\rm optimal} = \frac{1}{R}(o|L^{-1}|o)\,,
\end{align}
so we can use the single-shot variance $\sigma_O^2 \coloneqq (o|L^{-1}|o)$ as a quantitative measure of sample complexity for the observable $O$. If we want to quantify the performance of a scheme that uses a particular ensemble $\mathcal{U}$ without specifying an observable $O$, we have two options. A conservative `worst-case' approach is to consider the largest variance over all observables of sub-unit norm $(o|o) \leq 1$. We thus define the worst-case variance
\begin{align}
	\sigma_{\rm worst}^2 = \sup_{o : (o|o) \leq 1} (o|L^{-1}|o) = \|L^{-1}\|_\infty\,,
\end{align}
where $\|\,\cdot\,\|_\infty$ is the spectral norm (the largest absolute eigenvalue). 

Alternatively, to characterise the variance one would expect in estimating a typical observable, we can compute the average over a random ensemble of quadratic observables $O$, defined by some probability density function $p(O)$. Assuming invariance of this ensemble $p(V O V^\dagger) = p(O)$ under Gaussian rotations $V$, we obtain an `average case' quantifier of sample complexity
\begin{align}
	\sigma_{\rm avg}^2 = \mathbb{E}_{O \sim \mathcal{O}} (o|L^{-1}|o) = \frac{1}{N^2}\tr [L^{-1}]\,.
\end{align}
Using Chebychev's inequality, these metrics can be converted into bounds on the probability of our estimate deviating by more than some target error $\varepsilon$ using $R$ repetitions
\begin{align}
	\text{Pr}\big[|\mathbb{E}[\hat{\theta}_O - \text{Tr}[\rho_0 O]| > \varepsilon\big] \leq \frac{\sigma^2}{R \varepsilon^2},
\end{align}
where $\sigma^2 = \sigma^2_O$, $\sigma^2_{\rm worst}$, or $\sigma^2_{\rm avg}$ as appropriate.

\mysection{Minimum requirements \label{sec:Minimum}}

So far we have not specified anything about the particular ensemble of unitaries $\mathcal{U}$, other than that they are Gaussian of the form in \cref{eq:hop_ham_ferm_used}. Here we identify two necessary requirements of the ensemble $\mathcal{U}$ for our scheme to be successful.

\textit{(1) Dimension of linear maps.---}If we want to estimate a collection of $N_{\rm obs}$ linearly independent quadratic observables $\mathcal{O} = \{O_i\}_{i=1}^{N_{\rm obs}}$, then the column rank of the measurement map $F$ needs to be at least as large as $N_{\rm obs}$. If this were not the case, the measurement data $\hat{\vec z}$ would lie in a subspace of dimension $< N_{\rm obs}$, and we would not be able to find a left inverse $G$ such that $GF$ preserves all the desired observables. In other words, there would necessarily be some observables in the span of $\mathcal{O}$ that are not accessible in any of the measurement bases used in the scheme.

If we want to reconstruct the whole correlation matrix, this means we must have $S N_{\rm tot} \geq N^2$. This can be achieved either by taking a large number of ancilla sites $N_{\rm anc}  \sim N^2$, or by using many different rotations $S \sim N$. For our local scheme, we look to reconstruct correlations $C_{\vec i \vec j}$ for which $\vec i$ and $\vec j$ are separated by at most some constant range, which means we only need $SN_{\rm tot}  \sim N$.

\textit{(2) Sublattice symmetry.---}On a bipartite lattice without the quasiperiodic potential, the hopping Hamiltonian respects a sublattice symmetry $\Gamma \mathcal{H} \Gamma = - \mathcal{H}$, where $\Gamma = \bigl(\begin{smallmatrix}
	I_A & 0 \\ 0 & -I_B
\end{smallmatrix}\bigr)$ imparts a relative $\pi$ phase between the two sublattices $A$ and $B$. Under such a quench, the fermion creation operators evolve as $c_{\vec i}(t) = \sum_{\vec j} [e^{-i t \mathcal{H}}]_{\vec i \vec j} c_{\vec j}$. Because of the sublattice symmetry, the coefficients $[e^{-i t \mathcal{H}}]_{\vec i \vec j}$ have a particular structure: they are real for $\vec{i}$ and $\vec{j}$ on the same sublattice and purely imaginary otherwise. This makes it is impossible to measure the imaginary (real) part of any intra- (inter-)sublattice correlation function $C_{\rm \vec i \vec j}$, and the map $F$ will hence always be singular. (An analogous observation was made for hopping Hamiltonians in Ref.~\cite{gluza2021recovering}.)

We consider the quasiperiodic potential proposed in \cref{sec:setting} to be a particularly simple and experimentally-friendly way to break sublattice symmetry, and the ability to control both $h$ and $\phi$ also allows us to construct a suitably expressive ensemble of unitaries $\mathcal{U}$. However, other approaches are in principle feasible. For instance, sublattice symmetry can be broken by introducing time-dependence in the Hamiltonian, even if the instantaneous Hamiltonian is always sublattice symmetric~\cite{McGinley2018}.

\begin{figure*}
	\centering
	\includegraphics[width=0.98\linewidth]{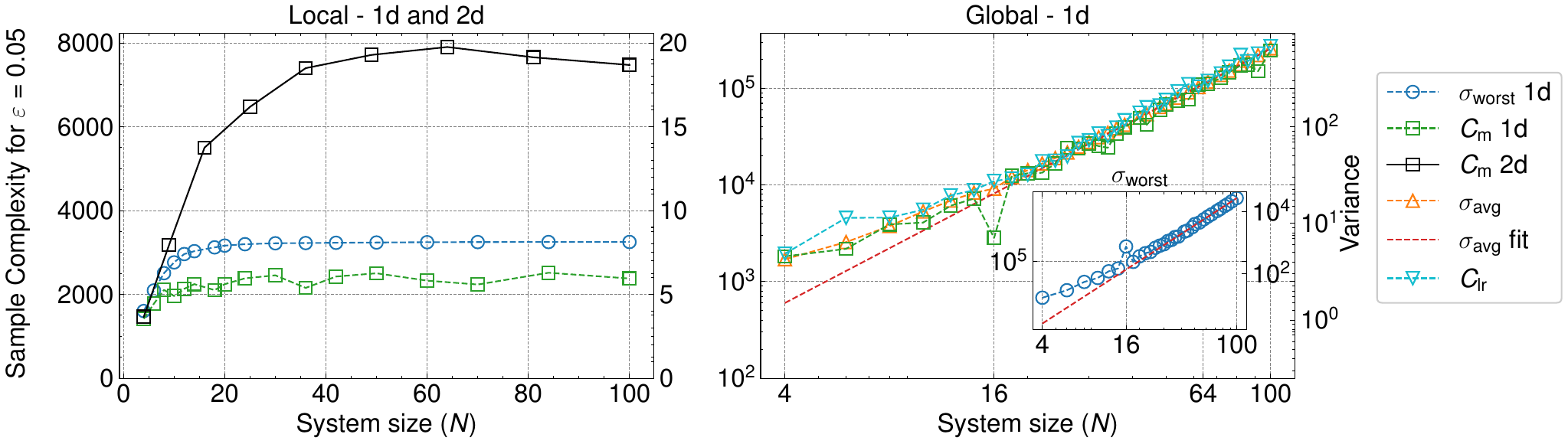}
	\caption{
    We display the sample complexity for various different cases (estimating worst, average and specific observables) and target accuracy of $\varepsilon=0.05$ (left y-axis) as well as the value of the variances (right y-axis).
		\textbf{Left:} We show the expected performance for estimating the worst-case local observable (blue) and the middle current (green (1d) / black (2d)) in 1d and 2d lattices. In both cases we set $h_{\rm max}=6$, $\theta_{\rm L}=2/3$ and singular value truncation at $\delta=10^{-3}$. In the 1d setting, we used the hyper-parameter settings of $t_{\rm max}=5$ and $S=400$. In the 2d setting, we estimated a single current observable with the hyper-parameter settings of $t_{\rm max}=1.5$ and $S=1000$. For building the inverse for systems with greater than $N=36$, we used the measurement-site truncation method described in \cref{app:truncation} with $\ell_{\rm out}=2$ to reduce the computational demand. 
		\textbf{Right:} Similar to the left panel but targeting global observables using $N^2+10N$ ancilla sites. We plot the average case sample complexity (orange) as well as the current in the middle (green) and a long range current where the distance is half the linear length of the one dimensional string (light blue). The inset shows the expected sample complexity for the worst case global observable. We do not randomize the phase or the strength of the laser and fix them at $\phi=0$ and $h=1$, respectively. Similarly, the laser angle and the evolution time are fixed at $\theta_{\rm L}\approx0.798$ and $t=0.75N$. The fit lines for the average- and worst-case follow asymptotic power law scaling with exponents $\alpha=1.88$ and $\alpha=2.88$, respectively.}
	\label{fig:protocol_performance}
\end{figure*}

\mysection{Results}\label{sec:results}

We now present numerical results to estimate the sample complexities of our methods in the two proposed schemes shown in \cref{fig:overview_protocols}: the ``local scheme"  to estimate local observables in 1d chains and 2d lattices and the ``global scheme" to estimating the arbitrary correlations in 1d chains. 

In the following, the parameters were not fully optimized with respect to sample complexity; rather, they were chosen to ensure feasible experimental settings characterzied by short evolution times, manageable ensemble sizes and practical sample complexities. We note that potential laser angles $\theta_{\rm L}$ close to 0, $\pi/4$ and $\pi/2$ radians should be avoided to prevent the near-chiral symmetry of the system. An ensemble of unitaries $\mathcal{U}$ of size $S$ was created by sampling $S$ tuples $(t,h, \phi)$ from the discretized grid of the space $(0,t_{\rm max}] \times (0, h_{\rm max}] \times (0,2\pi)$, where each dimension was split into 30 intervals. In each case, we give the specific parameter values in the caption of \cref{fig:protocol_performance}.

\mysubsection{Estimating local observables without ancillas}
In the left panel of \cref{fig:protocol_performance}, we show the scaling of the expected sample complexity (left y-axis) and the variance (right y-axis) when using the ``local scheme" without ancillas for systems of up to $N=100$ sites. In the 1d regime, we estimated both the worst-case local observable $\sigma_{\rm worst}$ and the current at the chain's midpoint $C_{\rm m}=\langle c_m\dagg c_{m+1}\rangle$, where $m$ is the midpoint index (or center point of a square layout). The worst-case sample complexity for estimating nearest-neighbor observables plateaus at approximately $R=3000$ samples, independent of the system size $N$, which indicates that the observable is indeed recovered locally. 
The sample complexity to estimate $C_{\rm m}$ has a lower plateau than that of the worst-case observable, but both are less than a factor of 10 away from the optimal shot noise limit of $1/\eps^2$ that one would achieve by directly measuring the observable, which shows that our protocol is fairly efficient.

In the 2d regime, we calculated the sample complexity for estimating the current in the middle of 2d square lattices. The expected sample complexity is around 7,500 measurements and independent of system size. The increased sample complexity compared to the 1d case is consistent with the increased connectivity and the correlation structure inherent in higher-dimensional lattices. Size independence again follows from finite velocity of the information under short time evolution with $t_{\rm max}<2$. We find it advantageous to set $t_{\rm max}<\ell=2$, as using $t_{\rm max}=2$ resulted in a significant increase in the variance for larger system sizes. 

Since the inverse reduces to the Moore-Penrose pseudoinversinverse, we employ the singular value decomposition and truncate the singular values at $\delta=10^{-3}$ prior to inverting the matrix. Careful selection of the truncation level $\delta$ is crucial for limiting the systematic error while balancing numerical stability and moderate sample complexities. However, since this can be done offline (and in principle even after the experiment has run), we do not expect this to be a practical issue.

\mysubsection{Estimating global observables with ancillas}To estimate global observables, we propose an alternative strategy, the ``global scheme" depicted in \cref{fig:overview_protocols} that expands the 1d target system into two dimensions by coupling it to an ancilla system of dimensions $L_x=N$ and $L_y=N+const.$, where $const.$ is a constant number of additional rows independent of system size. A single Hamiltonian governs the quench evolution, following the configuration shown on the right side of \cref{fig:overview_protocols}. Unlike in the local case, we randomize neither phase, nor strength of the potential laser, nor the evolution time.
Of course this can be added back in, but here we were aiming to showcase that this is not strictly required.
However, the evolution time needs to scale linearly to ensure the light cone of the quench Hamiltonian can propagate the information across the system.

The right panel of \cref{fig:protocol_performance} shows scaling behavior of the sample complexity to achieve a desired accuracy of $\varepsilon=0.05$ (left y-axis) expected variance (right y-axis) as one increases the system size $N$. The sample complexities for estimating the average- and worst-cases follow asymptotic power law scaling with respect to system size $N$ with exponents $\alpha<2$ and $\alpha<3$, respectively. Although estimating the worst-case observable for grids greater than $4\times 4$ demands an impractically high number of samples ($\gg10^5$ samples, as shown in the inset), a randomly chosen or average-case observable would require only $\sim10^5$ samples to reach an accuracy of $\varepsilon<0.05$ even for chains of 100 sites. The sample complexity for estimating a current in the middle of the system $C_{\rm m}$ is in line with the average case. Notably, a long-range current $C_{\rm lr}=\langle c_{\rm i}\dagg c_{\rm i+d}\pdagg\rangle$, such as one spanning a distance $d=N/2$, can be reconstructed with similar sample complexities to that of the middle current $C_{\rm m}$, measured between nearest neighbors ($d=1$). This finding suggests that our global estimation method is independent of the spatial locality of the observable.

\begin{figure*}
    \centering
    \includegraphics[width=0.9\linewidth]{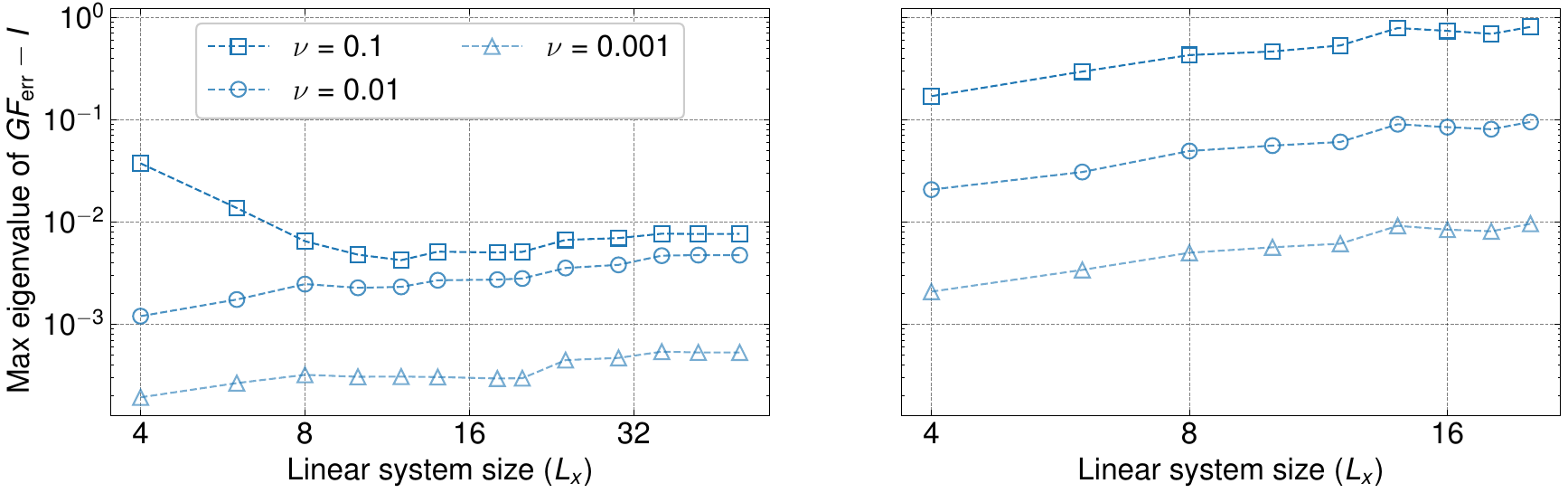}
    \caption{
    This shows the maximum eigenvalue of $G^{\rm opt}F_{\rm err}-\mathbb{I}$ for various different values of error $\nu$ for a perturbed forward map $F$ to get an error map $F_{\rm err}$ whereby the Hamiltonian has some random potential that has values sampled according to a standard distribution centered at zero with variance $\nu$. In both cases, we show the maximum across 50 random simulated error perturbations. \textbf{Left:} Local correlations estimation in a 1d chain without ancillas. \textbf{Right:} Global correlations estimation in a 1d chain with ancillas. Here, various combinations of the strength $h\in [0,5]$ and phase $\phi\in[0,\pi/2]$ of the Hamiltonian were tested with the best performing parameter combination selected for display. }
    \label{fig:error_sys_size_1d_global_local}
\end{figure*}
\mysubsection{Robustness to Errors}

We evaluate the error due to bias in the evolution Hamiltonian by adding a random potential to the Hamiltonian, sampled from a Gaussian distribution of variance $\nu$.
The maximal deviation in the correlation estimates is dictated by the largest eigenvalue of
\begin{equation}
    G^{\rm opt}F_{\rm err}-\mathbb{I}\,
\end{equation}
where $F_{\rm err}$ is the forward map with a biased (error) Hamiltonian. In \cref{fig:error_sys_size_1d_global_local} we show the maximum eigenvalue of $G^{\rm opt}F_{\rm err}-\mathbb{I}$ for several values of $\nu$ and for both schemes.
In the local scheme, which employs constant time evolution, the errors grow only very mildly with system size as the reconstruction is essentially local.
The global scheme instead displays a stronger sensitivity to tuning errors due to the long time evolution.

\mysection{Conclusions and Future Work}\label{sec:conclusion}
We present a practical and flexible framework for efficiently estimating  correlation matrices of quantum states prepared in neutral atom optical lattices using $10^3-10^5$ occupation measurements, featuring easily implemented quench Hamiltonians and classically efficient post-processing. The approach accommodates diverse laboratory configurations by combining or adapting the proposed schemes as needed.
Perhaps most interesting experimentally is our scheme to estimate local observables, as it features very competitive sample complexities of a few thousand shots and low evolution times of at most five hopping times.

In practice, we found that parameters have to be chosen judiciously to achieve optimal performance. However, calculating the expected sample complexity is efficient and can be performed on a laptop, which enables rapid optimization even for large systems.

Although we focus on bilinears, our method generalizes to higher-order observables (see \cref{app:higher_k_values}), extending previous non-interacting quench approaches~\cite{gluza2021recovering,denzler2024learning} to $k$-point functions.
The underlying enabling principle is that noninteracting evolution connects $k$-point correlators only to other $k$-point correlators, of which there are $N^k$ (or still linearly many if we restrict to local ones).

Our work lays a foundation for many promising areas of future development. The adaptation to continuous-space systems would extend its applicability beyond discrete lattice models. The exploration of time-dependent Hamiltonians in the quench evolution may offer additional advantages for noise robustness or sample efficiency. An investigation into the extension of using local kinetic measurements \cite{impertro2024local} rather than just the single-site measurements could be an interesting direction to further reduce the sample complexity. A comprehensive analysis of the protocol's robustness to errors to Hamiltonian errors or system noise would enable better parameter selection, further enhancing the method's practical utility.

\section*{Acknowledgements}
We are grateful to Bhavik Kumar for collaboration in a related project.
MM acknowledges support from Trinity College, Cambridge.
DM acknowledges financial support by the Novo Nordisk Foundation under grant numbers NNF22OC0071934 and NNF20OC0059939.

\appendix

\mysection{Estimating four-point correlators}\label{app:higher_k_values}
	
The scheme we describe in the main text can be straightforwardly generalized to estimate four-point (and in principle $k$-point) correlation functions, without assuming that the target state $\rho_0$ is Gaussian. For simplicity we assume that there are no ancillas here, though this can also be included if needed. In place of the correlation matrix $C_{\vec i \vec j}$, which we arranged as a $N^2$-component column vector contains all two-point correlators, we instead compile a column vector $|D_0)$ containing all two- and four-point correlation functions of the state $\rho_0$; explicitly, we have $|D) = |C_0) \oplus |D'_0)$, where $(\vec i \vec j \vec k \vec l|D'_0) = \text{Tr}[\rho_0 c^\dagger_{\vec i}c_{\vec j}c_{\vec k}^\dagger c_{\vec l}]$. Because the unitaries $e^{-i t_s H(h_s, \phi_s)}$ that we use in our protocol are Gaussian, they will transform $2k$-fermion operators to linear combinations of $2l$-fermion operators for all $l \leq k$. Thus, there will be a linear relationship between $|D_0)$ and the correlation vector for the rotated state $|D_s)$, defined the same way as $|D_0)$ with $\rho_s = U_s \rho U_s^\dagger$ in place of $\rho_0$. We write this as $|D_s) = U_s^{(4)}|D_0)$, where $U_s^{(4)}$ can be constructed from the matrix elements single-particle unitary $e^{-i t_s \mathcal{H}_s}$. The computational cost of constructing this map is still polynomial in $N$, since we are restricting ourselves to the space of $\leq 4$-fermion operators. 

When we measure the occupation numbers at all sites $\hat{\vec n} \in \{0,1\}^{N}$, we now construct a vector of not just the single-site occupations, but also two-site density-density correlations:
\begin{align}
    \hat{\vec n}^{(4)} = \Pi \big(\hat{\vec n} \oplus (\hat{\vec n} \otimes \hat{\vec n})\big)
\end{align}
i.e.~the components of $\hat{\vec n}^{(4)}$ are either of the form $\hat{n}_{\vec j}$ or $\hat{n}_{\vec j} \hat{n}_{\vec j'}$. The projector $\Pi$ removes any repeated elements, so there should be $N$ components for single-site occupations plus $N(N-1)/2$ two-site products. Conditioned on a particular value of $s$, the expectation value of this larger vector is a linear function of $|D_s)$
\begin{align}
    \mathbb{E}[\hat{\vec n}^{(4)} | s] = B|D_s) = F_s|D_0)
\end{align}
where $B$ is a linear map, and $F_s \coloneqq B U^{(4)}_s$. This is the analogue of \cref{eq:OccExpectation}. Once this is established, the post-processing continues as in the two-body case: From a given experimental run where we choose unitary $\hat{s}$ and observe occupation numbers $\hat{n}$, we create a vector $\hat{z}^{(4)} = \hat{\vec n}^{(4)}\otimes {\vec e}_s$. This gives us a linear relationship between the expectation value of $\hat{z}^{(4)}$ and the desired 4-point correlation functions
\begin{align}
    \mathbb{E}[\hat{\vec z}^{(4)}] &= F|D_0) & \text{where }F = \begin{pmatrix}
        p_1F_1 \\ \vdots \\ p_SF_S
    \end{pmatrix}
\end{align}
By computing a left inverse of $F$ (assuming one exists), we can infer four-point correlation functions.

\mysection{Variance and optimality of the estimator}\label{app:variance}

Suppose we use the scheme described in \cref{sec:methods} to estimate a particular quadratic observable $\braket{A} = (a|C_0)$, where $A = \sum_{\vec i \vec j} a_{\vec i \vec j} {c}^\dagger_{\vec i} {c}_{\vec j}^{\vphantom{\dagger}}$. For a given run of the experiment, we obtain measurements of the occupation numbers $\hat{\vec n} \in \{0,1\}^{N_{\rm tot}}$, and using the method described in the main text we construct an estimator of the correlation matrix $|\hat{Y}) = G(\hat{\vec z} - \vec d_{\rm anc})$, where $G$ is the left inverse of $F$.  The quantity $\hat{\theta}_A \coloneqq (a|\hat{Y})$ is then our desired estimator, $\mathbb{E}[\hat{\theta}_A] = \braket{A}$. As explained in the main text, depending on the dimensions of the linear map $F$, there might not be a unique choice of left inverse $G$, and thus, there will be different possible estimators. Our aim is to make a choice that minimizes the sample complexity of the protocol.

Conditioned on a particular choice of unitary $s$, our estimator $\hat{\theta}_A$ is a linear function of the observed measurement outcomes $\hat{\vec n}$, so we can write $\hat{\theta}_A|_s = \sum_{\vec j} b_{s, \vec j}\hat{n}_{\vec j} + \text{const.}$ for some set of coefficients $b_{s, \vec j}$, where the constant part is non-random and based on the starting ancilla state. Specifically, as a column vector we have $\vec{b} = G^T |a^*)$. The number of repetitions required to obtain an estimate of a desired accuracy can be determined by calculating the variance
\begin{equation}
\begin{aligned}
	\text{Var}[\hat{\theta}_A] &= \sum_s p_s \text{Var}[\hat{\theta}_A|_s]\\
    &= \sum_s p_s \sum_{\vec j, \vec j'} b_{s, \vec j}^* b_{s, \vec j'}^{\vphantom{*}}\big(\mathbb{E}\big[\hat{n}_{\vec j}\hat{n}_{\vec j'}|s\big] - \mathbb{E}\big[\hat{n}_{\vec j}|s\big]\mathbb{E}\big[\hat{n}_{\vec j'}|s\big]\big)
\end{aligned}
\end{equation}
where $\mathbb{E}[\hat{O}|s]$ is the expectation value of the observable $\hat{O}$ in the state $V_s(\rho_0 \otimes \rho_{\rm anc}) V_s^\dagger$ for $V_s=e^{-it_sH(h_s, \phi_s)}$. The right hand side of the above depends on the target state $\rho_0$, which we do not know in advance. As described in Ref.~\cite{Scott2006}, an appropriate solution is to eliminate this dependence by supposing that $\rho_0$ is drawn from some suitably structureless ensemble $\rho_0 \sim \mathcal{E}$, reflecting our lack of \textit{a priori} knowledge about the state in question. The precise choice of ensemble is not important, but we assume that the average state is $\mathbb{E}_{\rho_0 \sim \mathcal{E}}[\rho_0] = \pi_{\rm sys}$, where $\pi_{\rm sys} = I_{\rm sys}/d_{\rm sys}$ is the maximally mixed state of the system. For large enough system sizes, the typical variance of the estimator is then approximately
\begin{align}
\begin{aligned}
	\mathbb{E}_{\rho_0 \sim \mathcal{E}}\text{Var} [\hat{\theta}_A] 
    &\approx \sum_s p_s \sum_{\vec j, \vec j'} b_{s, \vec j}^* b_{s, \vec j'}\\
    &\times\Big(\text{Tr}\big[\rho_s \hat{n}_{\vec j}\hat{n}_{\vec j'}\big] - \text{Tr}\big[\rho_s\hat{n}_{\vec j}\big]\text{Tr}\big[\rho_s \hat{n}_{\vec j'}\big]\Big)
\end{aligned}
\end{align}
where $\rho_s \coloneqq V_s(\pi_{\rm sys}\otimes \rho_{\rm anc})V_s^\dagger $. (For instance, this can be explicitly verified when $\mathcal{E}$ is the Haar ensemble, by neglecting terms of order $d_{\rm sys}^{-1} = 2^{-N}$.) Assuming that $\rho_{\rm anc}$ is a Gaussian state, Wick's theorem can be applied to give
\begin{align}
	\mathbb{E}_{\rho_0 \sim \mathcal{E}}\text{Var} [\hat{\theta}_A] \approx \sum_s p_s  b_{s,\vec j}^*[W_s]_{\vec j \vec j'} b_{s, \vec{j}'} = \vec b^\dagger W \vec b
\end{align}
where
\begin{align}
    [W_s]_{\vec j \vec j'} &= \delta_{\vec j \vec j'}[C_s]_{\vec j \vec j'} - |[C_s]_{\vec j \vec j'}|^2
\end{align}
and
\begin{align}
	C_s & \coloneqq U_s^*\big((I/2)\oplus C_{\rm anc}\big)U_s^T .
\end{align}
	Here, we define the matrix $W = \oplus_{s=1}^S p_s W_s$, whose dimensions are $N_{\rm tot}S \times N_{\rm tot}S$. We look to find the \textit{optimal estimator}, namely the choice of $b_{s, \vec j}$ that minimizes the above variance, subject to the constraint that $\mathbb{E}[\hat{\theta}_A] = \braket{A}$. This is an optimization problem with a quadratic cost function and a linear constraint, which can be straightforwardly solved. This amounts to choosing the inverse
	\begin{align}
		G = (F^\dagger W^{-1} F)^{-1} F^\dagger W^{-1}
	\end{align}
	as claimed in the main text.

\mysection{Approximate local inverse}\label{app:truncation}
The measurement map $F$ defined in \cref{eq:z-expect} has $SN^3$ entries, which can require an excessive amount of memory to store. 
For example, for $S=1000$ and $N=121=11^2$, storing $F$ requires approximately 26 GB of memory.
As the system size increases further, storing $F$ becomes impractical.
In the local schemes, in which we aim to only recover local correlation functions, we only use constant time evolution, and thus we can instead define an approximate local version of $F$ that becomes independent of system size.

Concretely, we can compute the elements of $C_0$ individually by restricting the measurement data to a patch around the target sites, and then construct the inverse using only this localized data.
We define the neighborhood around a target site $\vec j$ as
\begin{equation}
    \K_\vec{j}^\ell=\{\vec i:\max(|i_x-j_x|,|i_y-j_y|)\leq\ell\}.
\end{equation}
The densities within this patch can only depend on its backward light cone (up to exponential tails).
Now let $\tilde F_{s,\vec j}^{(\ell_\mathrm{in},\ell_\mathrm{out})}$ be the map obtained from $F_s$ by keeping only the rows in $\K_\vec{j}^{\ell_\mathrm{out}}$ and the columns in $\K_\vec{j}^{\ell_\mathrm{in}}$.
The truncated maps are then concatenated as before to construct
\begin{equation}
    \tilde F_{\vec j}^\ell = 
    \mat{
        p_1\tilde F_{1,\vec j}^{(\ell_\mathrm{in},\ell_\mathrm{out})}\, &
        p_2\tilde F_{2,\vec j}^{(\ell_\mathrm{in},\ell_\mathrm{out})}\, &
        \cdots\, &
        p_S\tilde F_{S,\vec j}^{(\ell_\mathrm{in},\ell_\mathrm{out})}
    }^T.
\end{equation}

By construction, $\tilde F$ no longer depends on system size.
However, this procedure introduces very small singular values, which naively would lead one to conclude that the sample complexity becomes very large. 
Intuitively, these small singular values stem from exponential tails of the time evolution map.
To overcome this, we can additionally perform a singular value decomposition on $\tilde F$ and discard all singular values below some threshold $\delta\in(0,1]$.
This truncation introduces a small bias in the estimator, which can be controlled by $\delta$.
In our numerical exploration, we use $\delta=10^{-3}$, which leads to a bias substantially smaller than the target accuracy.
Overall, using $\tilde F$ for judiciously chosen $\ell_\mathrm{in}$, $\ell_\mathrm{out}$, and $\delta$ yields an estimation scheme that can easily be run on a laptop for arbitrary system sizes and whose error is only marginally increased compared to inverting $F$.

\bibliography{main}
\end{document}